\documentclass[aps,pra,twocolumn,showpacs]{revtex4}
\usepackage{graphicx}
\usepackage{amsfonts}
\usepackage{amssymb}
\usepackage{array}
\usepackage{amsmath}
\usepackage{color}
\usepackage{verbatim}
\newcommand{\ket}[1]{\left\vert#1\right\rangle}

\def\ket#1{|#1 \rangle}
\def\bracket#1#2{\langle #1|#2 \rangle}

\begin{document}
\title{Bell-state measurement and quantum teleportation using linear optics:
two-photon pairs, entangled coherent states, and hybrid entanglement}

\author{Seung-Woo Lee}
\thanks{Current Address: Department of Physics and Astronomy, Dartmouth College, 6127 Wilder Laboratory, Hanover, NH 03755, USA}
\affiliation{Center for Macroscopic Quantum Control,  Department of Physics and Astronomy, Seoul National University, Seoul, 151-742, Korea}

\author{Hyunseok Jeong}
\affiliation{Center for Macroscopic Quantum Control,  Department of Physics and Astronomy,
Seoul National University, Seoul, 151-742, Korea}

\date{\today}

\begin{abstract}
We review and compare Bell-state measurement and quantum teleportation schemes using linear optics with three different types of resources, {\it i.e.}, two-photon pairs, entangled coherent states and hybrid entangled states. Remarkably, perfect teleportation with linear optics is possible in principle based on a hybrid approach that combines two-photon pairs and entangled coherent states. It turns out that hybrid approach allows to perform Bell-state measurement and local Pauli operations in a deterministic way, while it requires photon number resolving measurements.
\end{abstract}

\maketitle
\section{introduction}

Quantum teleportation \cite{BennettPRL1993}, a task of transferring an unknown qubit between separated parties, is at the heart of various applications of quantum information processing. It particularly plays a crucial role in implementations of scalable optical quantum computation, as it enables one to perform sophisticated gate operations %necessary for quantum computation
using the gate teleportation protocol \cite{Gottesman1999}. The key element of the teleportation process is the Bell-state measurement, which discriminates between four two-mode entangled states known as Bell states
\begin{equation}
\label{eq:BellB}
\begin{aligned}
\ket{\Phi^\pm}=\frac{1}{\sqrt{2}}(\ket{0}\ket{0}\pm\ket{1}\ket{1})\\
\ket{\Psi^\pm}=\frac{1}{\sqrt{2}}(\ket{0}\ket{1}\pm\ket{1}\ket{0}).
\end{aligned}
\end{equation}
Another key element is to perform feed-forward transforms, $\hat{X}$ and $\hat{Z}$, on the output qubit to complete the whole teleportation process. Quantum teleportation has been experimentally demonstrated using optical systems \cite{exp1998}, and developed with various approaches.

A well known approach to optical quantum teleportation \cite{exp1998} employs single photon qubits, typically, with its polarization basis \{$|H\rangle$, $|V\rangle$\}, \cite{Knill2001,Kok07,Ralph10}. Alternatively, the vacuum and single photon, $|0\rangle$ and $|1\rangle$, may be used \cite{LundPRA2002}. In this type of approach, a two-photon pair, which can be generated using parametric down conversion, is required as the quantum channel. However, a deterministic Bell-state measurement cannot be performed using linear optics and photon detection: In fact, when single photon qubits are used, only two of the four Bell states can be identified and thus the success probability cannot exceed 50\% \cite{Lutkenhaus1999,Calsa2001}. In order to increase the success probability of a Bell measurement, a heavy increase of resources are required \cite{Knill2001,Kok07,Ralph10,Grice2011}.

Another approach using entangled coherent states has been studied with its remarkable merit \cite{vanEnkPRA2001,JeongPRA2001}.
This approach employs two coherent states, $\ket{\alpha}$ and $\ket{-\alpha}$ with amplitudes $\pm\alpha$ as a qubit basis. Van Enk and Hirota found that an entangled coherent state can be used as a quantum channel to teleport
a coherent-state qubit \cite{vanEnkPRA2001}. Jeong {\it et al.} showed that a nearly deterministic Bell state measurement can be realized for teleportation using an entangled coherent state as the quantum channel \cite{JeongPRA2001,Jeong2002QIC}.
It requires parity measurements to discriminate between the four Bell states, and its failure occurs when no photon is detected in both the detectors due to the nonzero overlap of $\bracket{0}{\pm\sqrt{2}\alpha}=e^{-\alpha^2}$.
Quantum computation schemes using coherent-state qubits  have also been explored along this line
\cite{JeongKim2001,Ralph2003,Lund2008}.
A critical problem in this approach is that one cannot easily perform the local unitary transforms required to finish the teleportation process. Due to non-orthogonality of two coherent states, $|\alpha\rangle$ and $|-\alpha\rangle$, the $\hat{Z}$ operation cannot be performed in a deterministic way. The teleportation between single photon and coherent state qubits was also studied \cite{Park2012}.

Recently, a new method that combines advantages of the two aforementioned approaches was proposed \cite{Lee2013}. It enables one to efficiently perform a near-deterministic quantum teleportation using hybrid entanglement. In this scheme, the orthonormal basis to define optical hybrid
qubits is $\big\{\ket{0_L}=\ket{+}\ket{\alpha},~\ket{1_L}=\ket{-}\ket{-\alpha}\big\}$ where $\ket{\pm}=(\ket{H}\pm\ket{V})/\sqrt{2}$. As the quantum channel, hybrid entanglement in the form of $|+\rangle|\alpha\rangle|+\rangle|\alpha\rangle+|-\rangle|-\alpha\rangle|-\rangle|-\alpha\rangle$ is required to implement this type of teleportation protocol. The Bell-state measurement in this context can be performed using two parallel Bell-state measurements, one for the polarization qubit part and the other for the coherent-state part, in a nearly deterministic way. Feed-forward logical unitary transforms, $\hat{X}$ and $\hat{Z}$, can be easily performed using both parts of the hybrid qubit to complete the whole teleportation process.

In this article, we review and compare linear optical quantum teleportation schemes with three different types of resources, {\it i.e.}, two-photon pairs, entangled coherent states and hybrid entangled states. It is shown that a nearly perfect teleportation can be implemented based on a hybrid approach, while other approaches have a weak point either in Bell-state measurement or feed-forward Pauli operations, both of which are essential tasks in quantum teleportation process. While an efficient generation of hybrid entanglement and a photon number resolving detection are still demanding, it may be challengeable in near future to experimentally demonstrate a nearly perfect teleportation using linear optics and hybrid entangled states along with current progress of optical quantum technologies.

\section{OPtical Bell measurements and quantum teleportation}

\subsection{Two photon pair}

Single photon is a primary candidate for optical quantum teleportation, which conveys information in its polarization degree of freedom \{$|H\rangle$, $|V\rangle$\} \cite{Knill2001} (or the vacuum and single photon basis, $|0\rangle$ and $|1\rangle$, \cite{LundPRA2002}). In the teleportation protocol, the sender (say Alice) and receiver (say Bob) are assumed to share a maximally entangled state, here chosen to be a two-photon pair {\it e.g.} $\ket{H}\ket{H}+\ket{V}\ket{V}$, which can be generated using parametric down conversion. Then, Alice performs Bell-state measurement between input qubit and one party of the channel state with her, aiming to discriminate four Bell states,
\begin{equation}
\label{eq:BellB}
\begin{aligned}
\ket{\Phi^\pm}=\frac{1}{\sqrt{2}}(\ket{H}\ket{H}\pm\ket{V}\ket{V})\\
\ket{\Psi^\pm}=\frac{1}{\sqrt{2}}(\ket{H}\ket{V}\pm\ket{V}\ket{H}).
\end{aligned}
\end{equation}
According to the measurement results, Bob applies Pauli operations, $\hat{X}$ or $\hat{Z}$, on his party of the channel to complete the teleportation process and retrieve the input state. Note that $\hat{X}$ and $\hat{Z}$ can be implemented deterministically in this approach, by polarization rotator and phase shift operation, respectively, whereas the success probability of Bell measurement cannot exceed 50\% as we shall see below.

In Fig.~\ref{fig1}(a), a Bell measurement scheme for single photon qubits ($\rm B_P$) is described by employing linear optic elements such as polarizing beam splitter (PBS), wave plates and photon detection. It has been known that this scheme allows to discriminate only two of the four Bell states \cite{Lutkenhaus1999,Calsa2001}. Suppose that $\ket{\Psi^+}$ or $\ket{\Psi^-}$ state enters into $\rm B_P$, at the first PBS two photons are separated into different modes resulting in one click from the upper two detectors and another from lower two. From all possible events of separated clicks, the two Bell basis $\ket{\Psi^+}$ and $\ket{\Psi^-}$ can be deterministically identified:
\begin{equation}
\begin{aligned}
{\rm (H,H)~or~(V,V)}&:&\ket{\Psi^-}\\
{\rm (H,V)~or~(V,H)}&:&\ket{\Psi^+}.
\end{aligned}
\end{equation}
On the other hand, for the state $\ket{\Phi^+}$ or $\ket{\Phi^-}$ two photons proceed together via the first PBS either way to upper or lower two detectors. As all possible results of clicks at the detectors from $\ket{\Phi^+}$ can be also yielded from $\ket{\Phi^-}$, it is impossible to discriminate these two states. Therefore, the overall success probability of Bell measurement (thus the success probability of quantum teleportation) is 50\%. Note that two identified Bell states out of four can be chosen by putting or removing appropriate wave plates at the input modes of the first PBS. For example, if we remove the $90^\circ$ wave plate in Fig.~\ref{fig1}(a), $\ket{\Phi^+}$ and $\ket{\Phi^-}$ states can be discriminated instead of $\ket{\Psi^-}$ and $\ket{\Psi^+}$.

In order to increase the success probability of quantum teleportation with single photon qubits, various methods have been developed, but they also encounter detrimental factors in a practical point of view. For example, a scheme of high-success teleporter requires a large number of modes prepared in single photon states \cite{Knill2001}. A scheme proposed by Grice \cite{Grice2011} can increase the success rate up to $1-1/2N$ by using $2^N-2$ number of ancillary photons and photon number resolving detection. Recently, Zaidi et al. \cite{Zaidi2013} suggested a method to improve the success probability up to 64.3 \% in dual-rail and to 62.5 \% in single rail approach in terms of inline squeezing operations accompanied by photon number resolving detectors.

\begin{figure}
\includegraphics[width=1\linewidth]{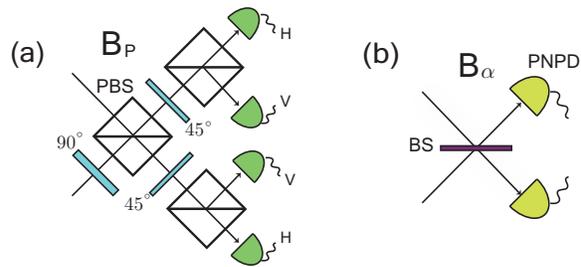}
\caption{Bell-state measurement schemes for (a) single photon polarization  qubit using polarizing Beam splitters (PBSs), wave plates, and on/off photo detectors, and for (b) coherent state qubits using beam splitter (BS) and photon number parity detectors (PNPDs).}\label{fig1}
\end{figure}

\subsection{Entangled coherent state}

Another well known approach employs coherent state qubits \cite{Sanders2012}, which have two coherent states, $\ket{\alpha}$ and $\ket{-\alpha}$ with amplitudes $\pm\alpha$ as a qubit basis. It was pointed out that nearly perfect Bell-state measurement can be performed using this approach \cite{JeongPRA2001,Jeong2002QIC}. As illustrated in Fig.~\ref{fig1}(b), the four Bell states of entangled coherent states,
\begin{equation}
\label{eq:BellB}
\begin{aligned}
\ket{\Phi^\pm}={\cal N^\pm}(\ket{\alpha}\ket{\alpha}\pm\ket{-\alpha}\ket{-\alpha})\\
\ket{\Psi^\pm}={\cal N^\pm}(\ket{\alpha}\ket{-\alpha}\pm\ket{-\alpha}\ket{\alpha}),
\end{aligned}
\end{equation}
with the normalization factor ${\cal N}_\pm=(2\pm2e^{-4|\alpha|^2})^{-1/2}$, can be identified using a 50:50 beam splitter (BS) and two photon number parity detectors (PNPDs) \cite{JeongPRA2001,Jeong2002QIC}. The Bell states after passing through the BS become
\begin{eqnarray}
\nonumber
\ket{\alpha}\ket{\alpha}\pm\ket{-\alpha}\ket{-\alpha}&\xrightarrow{\rm
BS}&\big(\ket{\sqrt{2}\alpha}\pm\ket{-\sqrt{2}\alpha}\big)\ket{0},\\
\nonumber
\ket{\alpha}\ket{-\alpha}\pm\ket{-\alpha}\ket{\alpha}&\xrightarrow{\rm
BS}&\ket{0}\big(\ket{\sqrt{2}\alpha}\pm\ket{-\sqrt{2}\alpha}\big),
\end{eqnarray}
where $\ket{\rm
even}\equiv{\cal N^+}\big(\ket{\sqrt{2}\alpha}+\ket{-\sqrt{2}\alpha}\big)$ and
$\ket{\rm odd}\equiv{\cal N^-}\big(\ket{\sqrt{2}\alpha}-\ket{-\sqrt{2}\alpha}\big)$ contain only even and odd number of photons, respectively. Therefore, from the results of two PNPDs, four Bell states can be discriminated:
\begin{equation}
\begin{aligned}
{\rm (even,0)}&:&\ket{\Phi^+},~~~
{\rm (0,even)}&:&\ket{\Psi^+}\\
{\rm (odd,0)}&:&\ket{\Phi^-},~~~
{\rm (0,odd)}&:&\ket{\Psi^-},
\end{aligned}
\end{equation}
where (even,0) indicates the detection of even number of photons at upper PNPD and no clicks at lower PNPD, and likewise for others. Due to the nonzero overlap of $\bracket{0}{\pm\sqrt{2}\alpha}=e^{-\alpha^2}$, the even number state $\ket{\rm even}$ possibly yields the case when no photon is detected in both detectors, which is counted as a failure of Bell measurement. As the amplitude $\alpha$ gets large, the failure rate of Bell measurement dramatically decreases.

However, this approach has a critical problem in feed-forward transforms, particularly with Pauli Z operation, which are required to finish the teleportation process at Bob's party. This difficulty in implementing $\hat{Z}$ operation is due to non-orthogonality of two coherent basis $\ket{\alpha}$ and $\ket{-\alpha}$, and may necessitate large cost of resources or heavy scheme complexity by repetition of gate teleportation \cite{Lund2008,Ralph2003}.

\subsection{Hybrid entanglement}

A hybrid approach has recently proposed by combining the two well known approaches, single photon and coherent state qubits \cite{Lee2013}. In this approach, the orthonormal basis to define optical hybrid qubits is
\begin{eqnarray}
\nonumber \big\{\ket{0_L}=\ket{+}\ket{\alpha},~~
\ket{1_L}=\ket{-}\ket{-\alpha}\big\},
\end{eqnarray}
where $\ket{\pm}=(\ket{H}\pm\ket{V})/\sqrt{2}$ and $\alpha$ is assumed to be real without loosing generality. As we shall see, this approach allows us to overcome particular weak points of previous approaches using single photon and coherent state qubits at the same time. Thus, quantum teleportation can be performed in a simple and near-deterministic manner using linear optics.

\begin{figure}
\includegraphics[width=0.9\linewidth]{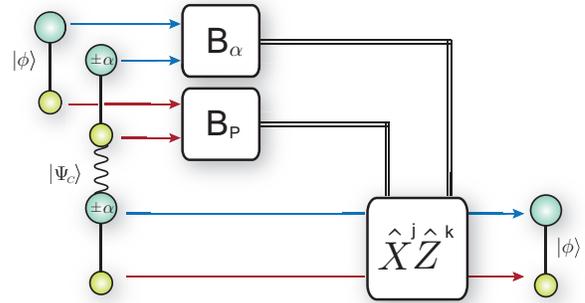}
\caption{Scheme for deterministic quantum teleportation in terms of linear optic elements and photon detection. An unknown hybrid qubit, $\ket{\phi}=a\ket{+}\ket{\alpha}+b\ket{-}\ket{-\alpha}$, is teleported via a hybrid entanglement channel $\ket{\Psi_C}$. Two Bell measurements $\rm B_{\alpha}$ and $\rm B_{P}$ are performed on coherent-state and single photon modes, respectively, between the input
qubit and one party of the channel state. Possible outcomes and corresponding feed-forward operations $\hat{X}^j\hat{Z}^k$ are presented in Table~\ref{tab:table1}. Its failure occurs only when both $\rm B_{\alpha}$ and $\rm B_{P}$ fail.
}\label{fig2}
\end{figure}

\begin{table}[b]
\caption{\label{tab:table1}{Feed-forwards dependent on $\rm B_{\alpha}$ and $\rm B_{P}$ results}}
\begin{tabular}{|c|c|c|}
\hline
$ \rm B_{\alpha}$ & $\rm B_{P}$ & Operations \\
\hline
(even, 0): j=0, k=0 & (H,H) or (V,V) &  \\
(odd, 0): j=0, k=1 & or (H,V) or (V,H) & \\
(0, even): j=1, k=0 & : flip k (0$\leftrightarrow$1) & $\hat{X}^j\hat{Z}^k$ \\
\cline{2-2}
(0, odd): j=1, k=1 & Otherwise: No flip & \\
\cline{1-2}
 & (H,V) or (V,H) : j=0, k=1 &\\
 (0, 0) & (H,H) or (V,V) : j=1, k=1 &\\
\cline{2-3}
 & Otherwise & Failure \\
\hline
\end{tabular}
%\end{ruledtabular}
\end{table}

The Pauli operation can be easily performed in this approach: $\hat{X}$ operation can be carried out by applying a bit flip operation on each of the two modes, i.e. implemented by a polarization rotator on the single photon mode and a $\pi$ phase shifter on the coherent state mode. The Pauli $Z$ operation, $\hat{Z}$, can be performed by applying a phase shift operation on the single-photon mode only, i.e. $\{\ket{+},~\ket{-}\} \rightarrow \{\ket{+},~e^{i\theta}\ket{-}\}$. Therefore, the feed-forward operations required to complete the teleportation process can be deterministically performed, which is a significant advantage over the highly nontrivial and resource demanding scheme used in coherent state approach \cite{Lund2008}.

The Bell-state measurement for an optical hybrid qubit can be performed by joint works of two Bell measurement schemes, $\rm B_{P}$ for single photon part and $\rm B_{\alpha}$ for coherent state part. Then, as we shall see below, it turns out that these hybrid approach enable to achieve a higher success probability of Bell-state measurement than the one obtained in either single approach.

Let us consider the details of hybrid teleportation scheme. Suppose that an unknown hybrid qubit, $\ket{\phi}=a\ket{0_L}+b\ket{1_L}$, and an entangled hybrid channel $\ket{\Psi_C}\propto \ket{0_L}_A\ket{0_L}_B+\ket{1_L}_A\ket{1_L}_B$ are prepared, where $A$ and $B$ denote the modes sent to Alice and Bob, respectively. The two units of Bell measurement, $\rm B_\alpha$ and $\rm B_{P}$, are performed in each physical mode between the input $\ket{\phi}$ and one party of the channel state $\ket{\Psi_C}$ as illustrated in Fig.~\ref{fig2}. From the results of the two Bell measurements, appropriate feed-forward transforms in the form of Pauli operations are determined as shown in the table~\ref{tab:table1}.

As performing $\rm B_{\alpha}$ on $\ket{\phi}$ and Alice's part of $\ket{\Psi_C}$, the coherent-state modes are mixed by the 50:50 BS and the total state evolves into
\begin{eqnarray}
\label{eq:afterBS}
&\ket{\phi}\ket{\Psi_C}&\xrightarrow{\rm BS}\\
\nonumber
&&\frac{1}{{\cal N^+}}{\ket{\rm even}}\ket{0}\Big(a\ket{+}\ket{+}\ket{0_L}_B+b\ket{-}\ket{-}\ket{1_L}_B\Big)
\\
\nonumber
&&+\frac{1}{{\cal N^-}}\ket{\rm odd}\ket{0}\Big(a\ket{+}\ket{+}\ket{0_L}_B-b\ket{-}\ket{-}\ket{1_L}_B\Big)\\
\nonumber
&&+\frac{1}{{\cal N^+}}{\ket{0}\ket{\rm even}}\Big(a\ket{+}\ket{-}\ket{1_L}_B+b\ket{-}\ket{+}\ket{0_L}_B\Big)\\
\nonumber
&&-\frac{1}{{\cal N^-}}\ket{0}\ket{\rm odd}\Big(a\ket{+}\ket{-}\ket{1_L}_B-b\ket{-}\ket{+}\ket{0_L}_B\Big).
\end{eqnarray}
Subsequent photon number parity measurements on output modes yield four possible outcomes for success events (even, 0), (odd, 0), (0, even), (0, odd), and a failure result, (0,0), when no photon is detected at both detectors.

For example, if the upper detector in $\rm B_\alpha$ detects an odd number of photons and the other does not, the outcome is (odd, 0) and we assign $j=0$ and $k=1$ as following the table~\ref{tab:table1}. As shown in Eq.~(\ref{eq:afterBS}), the remaining state is then $a\ket{+}\ket{+}\ket{0_L}_B-b\ket{-}\ket{-}\ket{1_L}_B$, which can be rewritten as
\begin{eqnarray}
\nonumber
&\Big(\ket{H}\ket{V}+\ket{V}\ket{H}\Big)\Big(a\ket{0_L}+b\ket{1_L}\Big)_B\\
&~~~~+\Big(\ket{H}\ket{H}+\ket{V}\ket{V}\Big)\Big(a\ket{0_L}-b\ket{1_L}\Big)_B.
\end{eqnarray}
Applying $\rm B_{P}$ measurements on single photon modes then verifies deterministically whether it projects onto $\ket{H}\ket{V}+\ket{V}\ket{H}$ or not. For example, if the outcome is $(H, V)$ or $(V, H)$, then the single photon modes projects onto $\ket{H}\ket{V}+\ket{V}\ket{H}$ so that the resulting state at Bob's party is $a\ket{0_L}+b\ket{1_L}$. In this case, as shown in the table~\ref{tab:table1}, we flip the assigned $k$ and thus finally have $j=0$ and $k=0$. Note that $(H, H)$ and $(V, V)$ clicks do not occur in this case. If $\rm B_{P}$ fails, the single photon modes projects onto $\ket{H}\ket{H}$ or $\ket{V}\ket{V}$ so that the state at Bob's party is now $a\ket{0_L}-b\ket{1_L}$. Then, the assigned $k$ remains unchanged and thus we have now $j=0$ and $k=1$. Therefore, the final feed-forward operation $\hat{X}^{j}\hat{Z}^{k}$ at Bob's party can restore the input state and complete the teleportation.

As an another example, let us consider the case when $\rm B_{\alpha}$ fails, i.e. no photon is detected at both PNPDs. As even number state $\ket{\rm even}$ only overlaps with the vacuum state, the corresponding state to the failure event of $\rm B_{\alpha}$ can be represented as
\begin{eqnarray}
\nonumber
&a\Big(\ket{+}\ket{+}\ket{0_L}_B+\ket{+}\ket{-}\ket{1_L}_B\Big)\\
\nonumber&~~~~~+b\Big(\ket{-}\ket{-}\ket{1_L}_B+\ket{-}\ket{+}\ket{0_L}_B\Big)\\
\nonumber
&=\big(\ket{H}\ket{V}+\ket{V}\ket{H}\big)\big(a\ket{0_L}-b\ket{1_L}\big)_B\\
\nonumber
&~~~-\big(\ket{H}\ket{V}-\ket{V}\ket{H}\big)\big(a\ket{1_L}-b\ket{0_L}\big)_B\\
\nonumber
&~~~+\big(\ket{H}\ket{H}+\ket{V}\ket{V}\big)\big(a\ket{0_L}+b\ket{1_L}\big)_B\\
&~~~+\big(\ket{H}\ket{H}-\ket{V}\ket{V}\big)\big(a\ket{0_L}-b\ket{1_L}\big)_B.
\end{eqnarray}
At the same time, if the outcome of $\rm B_{P}$ is (H,V) or (V,H), the single photon party is projected onto $\ket{H}\ket{V}+\ket{V}\ket{H}$ and $a\ket{0_L}-b\ket{1_L}$ is teleported to Bob. In this case, we assign $j=0$ and $k=1$ as following the Table.~\ref{tab:table1}. Likewise, for the outcomes (H,H) or (V,V), the state with Bob is $a\ket{1_L}-b\ket{0_L}$, and we assign $j=1$ and $k=1$. Finally, the input state can be retrieved by applying feed-forward operation $\hat{X}^{j}\hat{Z}^{k}$ on Bob's state.

Therefore, quantum teleportation can be successfully performed unless both $\rm B_\alpha$ and $\rm B_{P}$ fail. The success probability of hybrid quantum teleportation is thus given as
\begin{equation}
P_s=1-\frac{e^{-2\alpha^2}}{2},
\label{eq:fail}
\end{equation}
which significantly outperforms the previous schemes using two-photon pairs or entangled coherent states.

\section{Generation of resource states}

Schemes for generating two-photon pairs have been proposed through various types of entangled photon sources e.g. quantum dot \cite{Benson2000,Stace2003} and parametric down conversion \cite{Sliwa2003}, and have been experimentally achieved \cite{Kuzmich2003,Yamamoto2003}.

Entangled coherent states can be produced by splitting a superposed coherent state (SCS) with a beam splitter. In principle, an optical SCS can be generated in a deterministic way by using Kerr nonlinear effects \cite{Yurke86,Gerry99}, but it requires very high nonlinear strength that is not achievable within current technology.
Theoretical investigations of non-deterministic schemes based on
conditional measurements with squeezing operations  \cite{Dakna97,Lund2004,Marek2008,CLee2012, Lam2006a,Lam2006b} and weak nonlinear interactions \cite{JeongKerr2004,JeongKerr2005,He2009} have been made.
Experimental generations of SCSs so far have been demonstrated based on various non-deterministic schemes \cite{Neergaard2006,Ourjoumtsev2006,Ourjoumtsev2007,Takahashi2008,Sasaki2008,Ourjoumtsev2009,Gerrits2010,Neergaard2010}.  For example,
a scheme to produce a SCS by conditioning an input Fock state $\ket{n}$ on the outcome of homodyne detection was experimentally demonstrated to obtain a SCS of $\alpha\approx1.6$ \cite{Ourjoumtsev2007}.
Recently, the weakness of locally generated SCSs to losses, an obstacle for applications, was shown to be circumvented by an experimental generation of spatially separated entangled coherent states using a very lossy channel \cite{Ourjoumtsev2009}. A preparation of arbitrary squeezed vacuum and a squeezed single photon was demonstrated experimentally by photon-subtraction technique, which may be also used for engineering SCSs \cite{Neergaard2010}.

\begin{figure}
\includegraphics[width=0.9\linewidth]{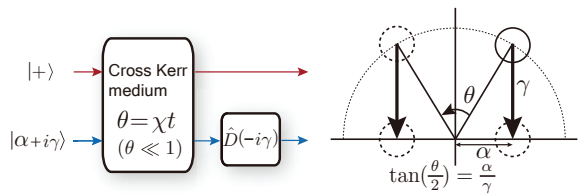}
\caption{
{\bf Scheme
to  produce a hybrid pair using an arbitrarily weak cross-Kerr
nonlinearity.}
A coherent state $\ket{\alpha+i\gamma}$ and a single photon in state
$\ket{+}=(|H\rangle+|V\rangle)/\sqrt{2}$ enter together into a weak cross-Kerr nonlinear medium
($\theta=\chi t \ll 1$). The values of $\theta$, $\gamma$, and
$\alpha$ are chosen to satisfy
$\gamma\tan(\frac{\theta}{2})=\alpha$. A displacement operation
$\hat{D}(-i\gamma)$ are performed on the coherent
state.}\label{generationHP}
\end{figure}

Hybrid entangled pairs ({\it e.g.} in the form of $|H\rangle |\alpha\rangle+|V\rangle|-\alpha\rangle$), can be also generated in principle using a cross-Kerr nonlinearity. The interaction Hamiltonian of the cross-Kerr nonlinearity is $H_{\chi}=\hbar\chi \hat{a}^{\dagger}\hat{a}\hat{b}^{\dagger}\hat{b}$ where $\hat{a}({\hat a}^{\dagger})$ and $\hat{b}({\hat b}^{\dagger})$ are the annihilation (creation) operators, and $\chi$ corresponds to the nonlinear strength. Suppose that a single photon in a diagonal state $\ket{+}=(\ket{H}+\ket{V})/\sqrt{2}$ interacts with a coherent state $\ket{\alpha}$ in a nonlinear medium. Using polarization beam splitters and the nonlinear interaction, a conditional phase shift can be performed so that
the output state then becomes entangled as $\ket{H}\ket{\alpha}+\ket{V}\ket{\alpha e^{i\theta}}$ where $\theta=\chi t$ with the interaction time $t$ \cite{Gerry1999}. Even though it is difficult to obtain large cross-Kerr nonlinearities such as $\chi t=\pi$, it is possible to generate hybrid pairs with an arbitrary weak nonlinearity following the ideas in Refs.~\cite{Nemoto2004,Jeong2005,Munro2005} as shown in Fig.~\ref{generationHP}. First, a coherent state as $\ket{\alpha+i\gamma}$ is required with a large real value $\gamma$, which interacts with the diagonal state $\ket{+}$ via a weak nonlinearity ($\theta\ll 1$). If a relation $\gamma\tan(\frac{\theta}{2}) = \alpha$ is satisfied, the output state becomes an entangled state $\ket{H}\ket{\alpha+i\gamma}+\ket{V}\ket{-\alpha+i\gamma}$. Finally, a hybrid pair can be obtained by applying the displacement operation $\hat{D}(-i\gamma)$ on the coherent state mode.

It is, however, still difficult to have an ideal cross-Kerr nonlinearity in optical single photon regime. It was pointed out that realistic multi-mode models could have limitations on the cross phase modulation due to spectral correlations between interacting fields in optical fibers that cause phase noises \cite{Shapiro2006,Shapiro2007}, and a similar result was obtained in  electromagnetically induced transparency media \cite{Banacloche2010}.

Nevertheless, it was demonstrated that these problems can be circumvented using an atomic V-type system and twin photons having temporal entanglement \cite{Koshino2009}. It was shown that high-fidelity and nonzero phase-shift are simultaneously obtainable in a multi-mode description of phase modulation using single photon and coherent state \cite{He2011}. Recently, schemes for large cross phase modulations were proposed by using a gradient echo memory \cite{Hosseini2011} and an atomic V-system \cite{Chudzick2012}. In this sense, it is highly expected that small-scale hybrid pairs, required for our scheme,
will be realized in the foreseeable future along with the recent progress in optical phase modulators \cite{Hosseini2011,Chudzick2012,Hwang2009,Hwang2011}.

\section{Conclusions}

We review quantum teleportation schemes of three different linear optical approaches using two-photon pairs, entangled coherent states and hybrid entangled states. We have particularly addressed the advantages of hybrid qubit approach that was recently proposed \cite{Lee2013}. In hybrid approach, the success probability of Bell-state measurement, an essential task for quantum teleportation, can be much larger than the one by single photon approach, 50\%. For example, $99\%$ success probability of teleportation can be achieved in hybrid scheme by encoding with $\alpha=1.4$. The Pauli operation used as a feed-forward task to complete the teleportation process, can be performed deterministically by simple linear optics, while in coherent state approach it is difficult to implement Pauli Z operation due to the non-orthogonality of coherent state basis. In addition, Ref.~\cite{Lee2013} also reported that hybrid approach is the most efficient way in the context of scalable quantum computation by considering fault-tolerant thresholds and resource requirements together.

Efficient schemes for generating hybrid entanglement between single photon and coherent state may be a challengeable next step for experimental realization. In principle, a hybrid entangled state can be generated via a weak cross-Kerr nonlinear interaction between a single photon and a strong coherent state \cite{Nemoto2004,Jeong2005,Munro2005}. Despite a limitation in optical fibers \cite{Shapiro2006,Shapiro2007}, recent studies have shown that a high-fidelity cross-Kerr nonlinearity can be obtained \cite{He2011,Hosseini2011,Chudzick2012}. Current progresses in photo detection techniques \cite{Eisaman2011} may enhance the possibility of efficient photon number resolving detection, which is another demanding task in hybrid approach. Therefore, an efficient quantum teleportation with hybrid entangle states is expected to be experimentally demonstrated in near future.

\acknowledgments
 
The authors are grateful to T. C. Ralph, W. J. Munro, B. C. Sanders, J. Lee, M. S. Kim, M. Paternostro, J. S. Neergaard-Nielsen, P. van Loock and X.-Q. Zhou for helpful discussions. We acknowledge financial support from National Research Foundation of Korea (NRF) grant funded by the Korean Government (No. 2010-0018295).

\end{document}